\documentclass[
    amsmath,
    amssymb,
    superscriptaddress,
    aps,
    prx,
    reprint,
    floatfix,
]{revtex4-2}

\usepackage{graphicx}
\usepackage{dcolumn}
\usepackage{glossaries}
\usepackage[version=3]{mhchem}
\usepackage[per-mode=symbol]{siunitx}
\usepackage[most]{tcolorbox}
\usepackage{xcolor}
\usepackage{hyperref}

\hypersetup{
    pdfauthor={Esmée Berger, Erik Fransson, Fredrik Eriksson, Eric Lindgren, Göran Wahnström, Thomas Holm Rod, Paul Erhart},
    pdftitle={Dynasor 2: From Simulation to Experiment Through Correlation Functions},
    unicode,
    colorlinks=true,
    urlcolor=black,
    linkcolor=black,
    citecolor=black
}

\newcommand{\e}{\text{e}}
\newcommand{\id}{\text{d}}
\renewcommand{\vec}[1]{\ensuremath\boldsymbol{#1}}

\newcommand{\gpumd}{\textsc{gpumd}}
\newcommand{\dynasor}{\textsc{dynasor}}
\newcommand{\ase}{\textsc{ase}}
\newcommand{\phonopy}{\textsc{phonopy}}

\newacronym{acf}{ACF}{autocorrelation function}
\newacronym{ins}{INS}{inelastic neutron scattering}
\newacronym{xrd}{XRD}{X-ray diffraction}
\newacronym{md}{MD}{molecular dynamics}
\newacronym{sed}{SED}{spectral energy density}
\newacronym{dho}{DHO}{damped harmonic oscillator}
\newacronym{mlip}{MLIP}{machine-learned interatomic potential}
\newacronym{nep}{NEP}{neuroevolution potential}
\newacronym{fcc}{FCC}{face-centered cubic}
\newacronym{sc}{SC}{simple cubic}

\DeclareSIUnit\angstrom{\text{Å}}
\DeclareSIUnit\rad{\text{rad}}
\newcommand{\hmn}[1]{
  \ensuremath{\begingroup\setupHMN #1\endgroup}%
}
\newcommand{\setupHMN}{%
  \doHMN{-}{\HMNoverline}%
  \doHMN{*}{\HMNminverse}%
  \doHMN{i}{\infty}
}
\newcommand{\doHMN}[2]{%
  \begingroup\lccode`~=`#1
  \lowercase{\endgroup\let~}#2%
  \mathcode`#1="8000
}
\newcommand{\HMNminverse}[1]{\frac{#1}{m}}
\newcommand{\HMNoverline}[1]{\mkern1mu\overline{\mkern-1mu#1\mkern-1mu}\mkern1mu}

\newcommand{\phys}{
    Department of Physics,
    Chalmers University of Technology,
    SE 412~96 Gothenburg, Sweden
}

\newcommand{\ess}{
    European Spallation Source ERIC,
    Copenhagen,
    Denmark
}

\begin{document}
\title{Dynasor 2:\texorpdfstring{\\}{} From Simulation to Experiment Through Correlation Functions}

\author{Esmée Berger}
\affiliation{\phys}
\author{Erik Fransson}
\affiliation{\phys}
\author{Fredrik Eriksson}
\affiliation{\phys}
\author{Eric Lindgren}
\affiliation{\phys}
\author{Göran Wahnström}
\affiliation{\phys}
\author{Thomas Holm Rod}
\affiliation{\ess}
\author{Paul Erhart}
\email{erhart@chalmers.se}
\affiliation{\phys}

\begin{abstract}
Correlation functions, such as static and dynamic structure factors, offer a versatile approach to analyzing atomic-scale structure and dynamics.
By having access to the full dynamics from atomistic simulations, they serve as valuable tools for understanding material behavior.
Experimentally, material properties are commonly probed through scattering measurements, which also provide access to static and dynamic structure factors. 
However, it is not trivial to decode these due to complex interactions between atomic motion and the probe. 
Atomistic simulations can help bridge this gap, allowing for detailed understanding of the underlying dynamics.
In this paper, we illustrate how correlation functions provide structural and dynamical insights from simulation and showcase the strong agreement with experiment.
To compute the correlation functions, we have updated the Python package \dynasor{} with a new interface and, importantly, added support for weighting the computed quantities with form factors or cross sections, facilitating direct comparison with probe-specific structure factors.
Additionally, we have incorporated the spectral energy density method, which offers an alternative view of the dispersion for crystalline systems, as well as functionality to project atomic dynamics onto phonon modes, enabling detailed analysis of specific phonon modes from atomistic simulation.
We illustrate the capabilities of \dynasor{} with diverse examples, ranging from liquid \ce{Ni3Al} to perovskites, and compare computed results with X-ray, electron and neutron scattering experiments. 
This highlights how computed correlation functions can not only agree well with experimental observations, but also provide deeper insight into the atomic-scale structure and dynamics of a material.
\end{abstract}

\maketitle

\section{Introduction}

The analysis of atomic-scale structure and dynamics is central in physics, chemistry, and materials science.
\Gls{md} simulations offer a detailed view into these atomic-level processes, providing trajectories of atomic positions and velocities from which structural and dynamical properties can be computed.

A powerful approach for extracting physically meaningful information from \gls{md} trajectories is through correlation functions, which capture dynamical properties in an interpretable form closely related to experimental observables.
Examples include static and dynamic structure factors, current correlations, \gls{sed}, and mode projection \glspl{acf}.
Unlike perturbative methods, \gls{md}-based correlation functions can inherently capture the full potential energy surface, enabling their application not only to crystalline but also to amorphous materials and liquids.

Experimentally, atomic structure and dynamics are often probed via scattering methods using electrons, neutrons, or X-rays.
These techniques produce correlation functions, in particular static and dynamic structure factors, but the raw data often requires theoretical modeling for interpretation due to probe-specific interactions and experimental limitations.
\Gls{md} simulations, with matching temporal and spatial scales, serve as an ideal complementary tool, enabling direct comparisons between computational predictions and experimental results.

A number of software packages have been developed to calculate scattering patterns and spectra from \gls{md} trajectories, including built-in modules within \gls{md} packages like \textsc{gromacs} \cite{BerVanVan95} and \textsc{lammps} \cite{Thompson2022}, as well as specialized tools like \textsc{nMoldyn} \cite{KneKeiKne95, RogMurHin03, HinPelSta12}, \textsc{mdanse} \cite{GorAouPel17}, \textsc{Sassena} \cite{LinSmi12}, \textsc{freud} \cite{freud2020} and \textsc{LiquidLib} \cite{WalJaiCaiZha}.
However, many of these packages specialize in particular scattering probes or offer limited post-processing capabilities.

\begin{figure*}
\centering
\includegraphics[width=\textwidth]{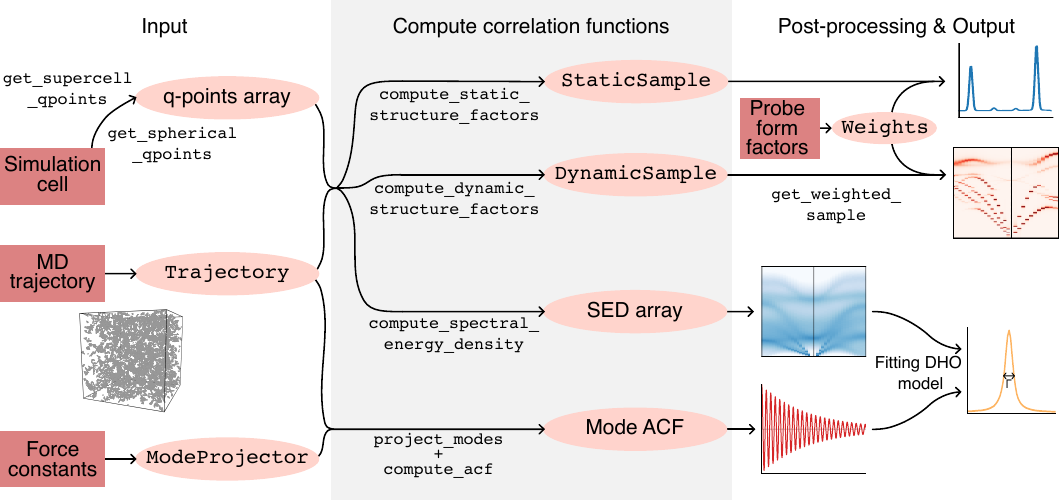}   
\caption{
    The \dynasor{} workflow typically consists of three parts: input, computation of the correlation functions, and handling of the output, which can include post-processing steps.
    The key component of the user input is a \gls{md} trajectory, which is parsed via the \ase{}- \cite{HjorthLarsen2017} or \textsc{mdanalysis}-based \cite{MichaudAgrawal2011, Gowers2016} \dynasor{} trajectory readers.
    In addition, a simulation cell or force constants are required, depending on which correlation functions are to be computed.
    The efficient computation of correlation functions is enabled by \textsc{numba} \cite{lam2015numba} and \textsc{numpy} \cite{harris2020array}.
    Post-processing includes weighting of static and dynamic samples with probe-specific weights (X-ray, neutron, electron form factors), and fitting correlation functions with the \gls{dho} model to extract phonon frequencies and lifetimes.
    User input and \dynasor{} data objects are indicated by rectangles and ovals, respectively.
    Key \dynasor{} functions are shown in lower-case monospaced font.
}
\label{fig:workflow}
\end{figure*}

In response, we here introduce version 2.X of the \dynasor{} package.
Compared to the 1.X release \cite{FraSlaErhWah2021}, the package has been substantially upgraded and extended to provide a fast and unified, Python-based interface capable of computing an even wider range of \glspl{acf} from \gls{md} trajectories (\autoref{fig:workflow}).
\dynasor{} now supports probe-specific weighting factors, including form factors and scattering lengths, enabling accurate and flexible computation of scattering patterns for multiple probe types.
Furthermore, advanced analysis such as mode projection is integrated, facilitating deeper insights into underlying dynamical processes.
In the next section we summarize the theoretical foundations underlying \dynasor{} before presenting several use cases that illustrate the various functionalities applied to X-ray, electron, and neutron probes.
These examples complement the extensive documentation available online \cite{dynasor_website}.

\section{Theoretical Background}

This section briefly summarizes the theoretical foundations of the methods implemented in \dynasor{}.
We first introduce correlation functions \cite{HanMcD2006, BooYip1980} as a general framework for analyzing materials dynamics, before outlining the \gls{sed} method \cite{Thomas2010}, which is suited for analyzing crystalline materials.
Next, we describe the mode projection technique \cite{SunSheAll2010, Zhang2014, Carreras2017, Rohskopf2022}, which provides detailed insights into mode dynamics in both ordered and disordered materials.
Finally, we summarize potential post-processing steps, including in particular the convolution of the dynamic structure factor with form factors to generate probe-specific spectra.

\subsection{Correlation Functions}\label{sec:theory-correlation-functions}

The partial intermediate scattering function $F_{AB}(\boldsymbol{q},t)$ is the spatial Fourier transform of the van Hove function, which describes the probability of finding a particle of type $A$ at position $\boldsymbol{r}$ at time $t$, when there is a particle of type $B$ at position $\boldsymbol{r}'$ at time $0$, or vice versa. 
$F_{AB}(\boldsymbol{q},t)$ can be expressed in terms of the spatially Fourier-transformed particle density $n_A(\boldsymbol{q},t)$ as
\begin{align*}
    F_{AB}(\boldsymbol{q},t) &= \frac{1}{\sqrt{N_A N_B}} \left< n_A(\boldsymbol{q},t)n_B(-\boldsymbol{q},0) \right>.
\end{align*}
Here, $\left < \ldots \right >$ indicates an ensemble average, or a time average if the system is ergodic, and $N_x$ is the number of particles of type $x$. 
For a \gls{md} simulation with classical dynamics, the particle density can be described by $n_A(\boldsymbol{q},t) = \sum_{i\in A}^{N_A} \e^{i\boldsymbol{q} \cdot \boldsymbol{r}_i(t)}$, such that 
\begin{equation*}
    F_{AB}(\boldsymbol{q},t) = \frac{1}{\sqrt{N_A N_B}}\sum_{i\in A}^{N_A} \sum_{j\in B}^{N_B} \left< \exp \left[i\boldsymbol{q}\cdot(\boldsymbol{r}_i(t)-\boldsymbol{r}_j(0))\right] \right>.
\end{equation*}

The total intermediate scattering function is obtained as a weighted sum over the partial ones, e.g., 
\begin{align*}
    F(\boldsymbol{q},t) 
    ={}&\frac{N_A}{N} F_{AA}(\boldsymbol{q},t) + \frac{N_B}{N} F_{BB}(\boldsymbol{q},t) \\ &+  \frac{2\sqrt{N_A N_B}}{N} F_{AB}(\boldsymbol{q},t)
\end{align*}
for a two-component system, where $N=N_A+N_B$ denotes the total number of atoms. 
In \dynasor{} we introduce new variables $\mathcal{F}_{AB}(\boldsymbol{q},t)$ for convenience. 
These can be viewed as relabeled partial intermediate scattering functions, which collect everything related to the types $A$ and $B$ in separate variables such that the sum simplifies to $F(\boldsymbol{q},t) = \mathcal{F}_{AA}(\boldsymbol{q},t) + \mathcal{F}_{AB}(\boldsymbol{q},t) + \mathcal{F}_{BB}(\boldsymbol{q},t)$. 
This means that the \dynasor{} partial intermediate scattering functions are
\begin{align*}
    \mathcal{F}_{xy}(\boldsymbol{q},t) &= \frac{(2-\delta_{xy})}{N}\sqrt{N_x N_y} F_{xy}(\boldsymbol{q},t) \\ &= \frac{(2-\delta_{xy})}{N}\sum_{i\in x}^{N_x} \sum_{j\in y}^{N_y} \left< \exp \left[i\boldsymbol{q}\cdot(\boldsymbol{r}_i(t)-\boldsymbol{r}_j(0))\right] \right>,
\end{align*}
where $\delta_{xy}$ is a Kronecker delta.

Furthermore, the total intermediate scattering function $F(\boldsymbol{q}, t)$ described above, which is the Fourier transform of the probability density for finding \emph{the same or any other} particle, is commonly referred to as the coherent intermediate scattering function $F_{\text{coh}}(\boldsymbol{q},t)$.
An important subset of this is the incoherent (or self) part, $F_{\text{incoh}}(\boldsymbol{q},t)$, which is the Fourier transform of the probability density of finding \emph{the same} particle at position $\boldsymbol{r}$ at time $t$ as was found at position $\boldsymbol{r}'$ at time 0, i.e.,
\begin{align*}
    F_{\text{incoh}}(\boldsymbol{q},t) &= 
    \sum_{A\in\text{types}}^{N_\text{types}} \mathcal{F}_{AA}^{\text{incoh}}(\boldsymbol{q},t) 
    \\ 
    &= \frac{1}{N}\sum_{A\in\text{types}}^{N_\text{types}}\sum_{i\in A}^{N_A}  \left< \exp \left[i\boldsymbol{q}\cdot(\boldsymbol{r}_i(t)-\boldsymbol{r}_i(0))\right] \right> \\ 
    &= \frac{1}{N}\sum_{i}^{N}  \left< \exp \left[i\boldsymbol{q}\cdot(\boldsymbol{r}_i(t)-\boldsymbol{r}_i(0))\right] \right>.
\end{align*}

Fourier transforming the partial intermediate scattering function in time yields the partial dynamic structure factor
\begin{equation*}
    S_{AB}(\boldsymbol{q},\omega) = \int _{-\infty} ^\infty \mathcal{F}_{AB}(\boldsymbol{q},t) \, e^{-iwt} \mathrm{d}t.
\end{equation*}
From the peaks in $S_{AB}(\boldsymbol{q},\omega)$ we can learn which modes involve interactions between particles of types $A$ and $B$, and, e.g., distinguish between diffusional motion at low frequencies and vibrational motion at higher frequencies.
The broadening of these peaks gives information about the phonon lifetimes, which we return to in \autoref{sec:post-processing}, and the overall appearance of them can indicate whether the motion is localized or not.

Analogously to the intermediate scattering function, current correlations can be defined in terms of spatially Fourier-transformed current densities, $\boldsymbol{j}(\boldsymbol{q},t) = \sum_{i}^{N} \boldsymbol{v}_i(t) \e^{i\boldsymbol{q} \cdot \boldsymbol{r}_i(t)}$, as
\begin{align*}
C(\boldsymbol{q},t) =&{} \frac{1}{N} \left< \boldsymbol{j}(\boldsymbol{q},t)\cdot\boldsymbol{j}(-\boldsymbol{q},0) \right>\\
    =&{} \underbrace{\frac{1}{N} \left< \boldsymbol{j}_L(\boldsymbol{q},t)\cdot\boldsymbol{j}_L(-\boldsymbol{q},0) \right>}_{C_L(\boldsymbol{q},t)} + \\
    &{} \underbrace{\frac{1}{N} \left< \boldsymbol{j}_T(\boldsymbol{q},t)\cdot\boldsymbol{j}_T(-\boldsymbol{q},0) \right>}_{C_T(\boldsymbol{q},t)}.
\end{align*}
Here, we have used that $\boldsymbol{j}(\boldsymbol{q},t) = \boldsymbol{j}_L(\boldsymbol{q},t) + \boldsymbol{j}_T(\boldsymbol{q},t)$, where the longitudinal current $\boldsymbol{j}_L(\boldsymbol{q},t)$ is in the direction of $\boldsymbol{q}$ and the transverse current $\boldsymbol{j}_T(\boldsymbol{q},t)$ is perpendicular to $\boldsymbol{q}$. 
As with the intermediate scattering function, the total current correlation $C(\boldsymbol{q},t)$ is related to the corresponding partials, $C_{AB}(\boldsymbol{q},t)$, via a direct sum.
Fourier transforming $C(\boldsymbol{q},t)$ yields the frequency-dependent current correlation $C(\boldsymbol{q},\omega)$, analogously to the relation between the intermediate scattering function and the dynamic structure factor.
Due to the particle density and current being related via $\dot{n}(\boldsymbol{q},t) = i\boldsymbol{q}\cdot\boldsymbol{j}(\boldsymbol{q},t)$, current correlations relate to the dynamic structure factor through 
\begin{equation}\label{eq:relation-S-Cl}
    \omega^2S(\boldsymbol{q},\omega) = q^2C_L(\boldsymbol{q},\omega),
\end{equation}
meaning that some features can be easier to resolve in either of these correlation functions.
Furthermore, current correlations can be interpreted as spatially-dependent generalizations of the velocity autocorrelation function. 
A closely related concept is the phonon \gls{sed} \cite{Thomas2010}, which is a measure of the distribution of kinetic energy in the $(\boldsymbol{q}, \omega)$-plane.

\subsection{Spectral Energy Density}

In systems where the atoms vibrate around equilibrium positions, as is the case in many solids, the dynamics can also be analyzed using the phonon \gls{sed} method.  
Here, we briefly present an overview of the theory underpinning this method as implemented in \dynasor{}.
For a more detailed description see, e.g., Ref.~\citenum{Thomas2010}.  

In the \gls{sed} approach, the starting point is the perfect crystalline lattice, with each atom $i$ in its equilibrium position $\boldsymbol{r}^0_i$.
By replacing the atomic index $i$ with unit cell index $\boldsymbol{l}$ and basis site index $s$, this yields the lattice current
\begin{equation*}
        \vec{j}^0(\boldsymbol{q}, t) = \sum_s \vec{j}_s^0(\boldsymbol{q}, t) = 
        \sum_s\sum_{\boldsymbol{l}} \vec{v}_s(\boldsymbol{l},t) \e^{i \vec{q} \cdot \vec{r}^0(\boldsymbol{l})},
\end{equation*}
where $\boldsymbol{j}_s^0(\boldsymbol{q}, t)$ is the partial lattice current for the atoms at basis site $s$, and $\boldsymbol{r}^0(\boldsymbol{l})$ is the position of unit cell $\boldsymbol{l}$.
The corresponding partial lattice current correlation function is 
\begin{equation*}
    K_s(\boldsymbol{q}, t) = \frac{1}{N_\text{unit}}\frac{m_s}{2}\left< \vec{j}_s^0(\boldsymbol{q}, t) \cdot \vec{j}_s^0(-\boldsymbol{q}, 0) \right>, 
\end{equation*}
where $N_\text{unit}=N/N_\text{basis}$ is the number of unit cells, $N_\text{basis}$ is the number of basis atoms in the primitive cell, and $m_s$ is the mass of basis atom $s$. Summing over all basis atoms, the total \gls{sed} is 
\begin{equation*}
    K(\boldsymbol{q},\omega) = \sum_s K_s(\boldsymbol{q}, \omega).
\end{equation*}
In comparison to the current correlations $C(\boldsymbol{q},\omega)$ from the previous section, we see that $K(\boldsymbol{q},\omega)$ can be thought of as a mass-weighted total lattice current correlation. Furthermore, we can note that the \gls{sed} integrates to the total kinetic energy of the system  
\begin{equation*}
    \int K(\boldsymbol{q},\omega) \id\omega = N_\text{basis}\frac{3}{2}k_\text{B} T, 
\end{equation*}
and that it is possible to decompose it spatially as $K(\boldsymbol{q},\omega) = K_x(\boldsymbol{q},\omega) + K_y(\boldsymbol{q},\omega) + K_z(\boldsymbol{q},\omega)$, similarly to the decomposition of $C(\boldsymbol{q},\omega)$ into transversal and longitudinal components. 
Despite the \gls{sed} method being limited to crystalline systems, whereas the current correlations $C(\boldsymbol{q},\omega)$ can be computed for any system, \gls{sed} can provide useful insights as it offers an alternative view of the phonon dispersion, with each mode having the same spectral weight.

\subsection{Mode Projection}

To gain further microscopic insight into the dynamics from a \gls{md} simulation, the motions can be interpreted in terms of the dynamics of phonon modes.
This technique is often called mode projection or normal mode decomposition, and it allows for understanding features in the scattering spectra in terms of correlated motion, which, in principle, can be linked to structural or electronic properties of the material.
In particular, mode projection links atomistic simulations to the theoretical framework of lattice dynamics used to study vibrations in crystals, and allows for extracting phonon properties with the full anharmonicity included in the \gls{md} simulation.
Ideally a basis (phonon eigenmodes) should be chosen to aid physical understanding, which could for example be the zero-Kelvin modes from the harmonic approximation, modes from a more symmetrical phase of the material of interest, or modes from ideal versions of subsystems in the material.

Several tools exist to perform mode projection, such as \textsc{dynaphopy} \cite{Carreras2017} and \textsc{modecode} \cite{Rohskopf2022}.
In the \dynasor{} implementation we emphasize the ease of use via the Python interface, as well as the smooth integration with other parts of the \dynasor{} package.

In the remainder of this section, we provide a brief introduction to the theory behind mode projection in the framework of lattice dynamics. 
For a more extensive description, see, e.g., Refs.~\citenum{SunSheAll2010, Zhang2014, Carreras2017, Rohskopf2022}.

Consider the potential part of the harmonic Hamiltonian describing vibrations in a harmonic crystal
\begin{equation*}
    U = \frac{1}{2}\sum_{\alpha\beta}\sum_{\boldsymbol{l}\boldsymbol{l'}}\sum_{ss'} \Phi_{ss'}^{\alpha\beta}(\boldsymbol{l},\boldsymbol{l'}) u_s^\alpha(\boldsymbol{l})u_{s'}^\beta(\boldsymbol{l'}).
\end{equation*}
Here, $u^\alpha_s(\boldsymbol{l})$ is the displacement in direction $\alpha$ of the atom at basis site $s$ in the unit cell $\boldsymbol{l}$, and $\Phi$ are harmonic force constants.
The Hamiltonian can alternatively be expressed in terms of mode coordinates $Q_b(\boldsymbol{q})$ for band $b$ at $q$-point $\boldsymbol{q}$ as
\begin{equation*}
    U = \frac{1}{2}\sum_{\boldsymbol{q}b}\omega^2_b(\boldsymbol{q}) Q_b(\boldsymbol{q}) Q_b(-\boldsymbol{q})
\end{equation*}
by transforming the displacements $u$ to mode coordinates $Q$ using
\begin{equation}
\begin{split}
    \label{eq:u_from_Q}
    u_s^\alpha(\boldsymbol{l})  &= \frac{1}{\sqrt{N_\text{unit}}}\sum_{\boldsymbol{q}b} \frac{1}{\sqrt{m_s}}Q_b(\boldsymbol{q}) W_{sb}^\alpha(\boldsymbol{q}) e^{i\boldsymbol{q} \cdot \boldsymbol{r}_s(\boldsymbol{l})}\\
    &= \sum_{\boldsymbol{q}b}X_{sb}^\alpha(\boldsymbol{l},\boldsymbol{q})Q_b(\boldsymbol{q}) 
    ,
\end{split}
\end{equation}
where $\boldsymbol{r}_s(\boldsymbol{l})$ is the position of the atom at basis site $s$ in unit cell $\boldsymbol{l}$, $X$ are eigenmodes, and $W$ polarization vectors.
The polarization vectors are defined via the dynamical matrix
\begin{align*}
    \label{eq:dynamical_matrix}
    D_{ss'}^{\alpha\beta}(\boldsymbol{q}) &= \frac{1}{\sqrt{m_s m_{s'}}}\sum_{\boldsymbol{l}}\Phi_{ss'}^{\alpha\beta}(0, \boldsymbol{l}) \e^{-i \boldsymbol{q}\cdot (\boldsymbol{r}_s(0)-\boldsymbol{r}_{s'}(\boldsymbol{l})) } \\ &= \sum_{b} W_{s b}^{\alpha}(\boldsymbol{q}) \omega_{b}^2(\boldsymbol{q}) W_{s' b}^{\beta}(-\boldsymbol{q}).
\end{align*}
The new mode coordinates are in the form of lattice waves
\begin{equation}
    \label{eq:Q_from_u}
    Q_b(\boldsymbol{q}) = \frac{1}{\sqrt{N_\text{unit}}} \sum_{s \alpha \boldsymbol{l}} \sqrt{m_s} u_s^\alpha(\boldsymbol{l}) W_{s b}^{\alpha}(-\boldsymbol{q})\e^{-i\boldsymbol{q}\cdot \boldsymbol{r}_s(\boldsymbol{l})}.
\end{equation}
Similarly, we can define the mode momenta $P_b(\boldsymbol{q})=\dot{Q}_b(-\boldsymbol{q})$ and the mode force $\dot{P}_b(\boldsymbol{q})$.

The normal modes are typically constructed from the harmonic force constants $\Phi$, but in principle any informed choice can be used.
With the transformations in Eq.~\eqref{eq:u_from_Q} and \eqref{eq:Q_from_u}, it is thus possible to freely change coordinates back and forth between the atomic displacements $u$ and the phonon coordinates $Q$.
This allows for the projection of \gls{md} trajectories, containing $u(t)$ (for all $s,\alpha,\boldsymbol{l}$), onto the eigenmodes to obtain $Q(t)$ (for all $b, \boldsymbol{q}$) in order to access a description of the dynamics in terms of phonon modes.
By analyzing the \glspl{acf} of the time-dependent mode coordinates, $ F^Q(t) = \left < Q(0) Q^*(t) \right >$, mode momenta, and their Fourier transforms (analogous to $F(\boldsymbol{q},t)$, $S(\boldsymbol{q}, \omega)$, $C(\boldsymbol{q}, t)$, and $C(\boldsymbol{q}, \omega)$ for the atomic motion described in \autoref{sec:theory-correlation-functions}) the renormalized frequencies and lifetimes of the modes can be extracted using \gls{dho} fitting \cite{SunSheAll2010}, which is one of the ways to post-process the computed correlation functions.
Notably, mode projection is closely related to \gls{sed}, as summing the velocity power spectra over band indices for a given $q$-point would contain the same information as the \gls{sed} for that $q$-point \cite{Thomas2010}.
In turn, as discussed previously, \gls{sed} is closely related to current correlations, which themselves are related to dynamic structure factor Eq.~\eqref{eq:relation-S-Cl}.

Importantly, phonon mode projection can also be used to gain insight into the structure of the system.
For example, in systems undergoing phase transitions driven by soft phonon modes, the mode coordinates $Q$ can serve as order parameters to understand the phase transitions better \cite{Rudin2018, Ouyang2021, FraRosEriRahTadErh23}.

\subsection{Post-Processing}\label{sec:post-processing}

By post-processing the computed correlation functions, additional information about the structure and dynamics can be obtained. 
There are several different tools in \dynasor{} that facilitate post-processing of the correlation functions, e.g., weighting, which is crucial for comparison to experiment, \gls{dho} fitting to extract phonon frequencies and lifetimes, and spherical averaging in $\boldsymbol{q}$, relevant for, e.g., liquid systems.

\subsubsection{Weighting}

The total dynamic structure factor $S(\boldsymbol{q},\omega)$ is obtained through a weighted sum of the partial structure factors
\begin{equation*}
    S(\boldsymbol{q},\omega) = \sum_A\sum_B w_A(\boldsymbol{q})w_B(\boldsymbol{q})S_{AB}(\boldsymbol{q},\omega),
\end{equation*}
where the weights $w_\alpha(\boldsymbol{q})$ can, e.g., be probe-specific form factors or scattering lengths, to indicate the interaction strengths between the probe and the atom types, or, e.g., atom-specific masses or charges, which can be useful to investigate acoustic or optical modes, respectively.
Computationally, when the weights $w_{\alpha}(\boldsymbol{q}) = w_{\beta}(\boldsymbol{q}) = 1$, the total raw dynamic structure factor is obtained, containing information about the full dynamics of the system. 
Experimentally, the dynamic structure factor is commonly probed in inelastic scattering experiments, whereas the static structure factor $S(\boldsymbol{q})$, given by the intermediate scattering function at $t=0$, is probed in diffraction experiments. 
By using probe-specific weights in \dynasor{}, the computed structure factors can be directly compared to experimentally measured ones.
For X-rays and electrons the weights are so-called form factors, which depend on $|\boldsymbol{q}|$, whereas the weights for neutrons are so-called neutron scattering lengths. 
When comparing to experiment in the examples in this paper we use the direct support for weighting implemented in \dynasor{}, with the weights being X-ray form factors from \cite{Waasmaier1995, itcc2006}, neutron scattering lengths from \cite{NISTneutrons}, and electron form factors from \cite{Peng1996atoms, Peng1998ions}.

\subsubsection{Damped Harmonic Oscillator Fitting}

To extract phonon frequencies $\omega_0$ and lifetimes $\tau$, dynamic structure factors, current correlations, \glspl{sed}, or mode coordinate \glspl{acf} can be fitted using the analytical expression for a \gls{dho}.
This can be done by modeling, e.g., the particle density $n(\boldsymbol{q}, t)$ as a \gls{dho} subject to a stochastic force, so the relevant equation of motion is  
\begin{equation*}
    \ddot{x}(t) + \Gamma\dot{x}(t)+\omega_0^2x(t) = f(t),
\end{equation*}
where $\omega_0$ is the natural angular frequency of the oscillator and $\Gamma = 2/\tau$ is the damping constant. 
Assuming that the stochastic force, $f(t)$, is white noise, the corresponding positional autocorrelation reads
\begin{equation*}
    \ddot{F}^\text{DHO}(t) + \Gamma\dot{F}^\text{DHO}(t)+\omega_0^2F^\text{DHO}(t) = 0.
\end{equation*}
Setting $F^\text{DHO}(0)=A$ and $\dot{F}^\text{DHO}(0)=0$ yields the solutions
\begin{equation*}
F^\text{DHO}(t) =
    \begin{cases}
     A\e^{-t/\tau}(\cos{\omega_e t + \frac{1}{\tau\omega_e}\sin{\omega_e t}}), & \omega_0>\frac{1}{\tau} \\
     A\e^{- t/\tau}(\cosh{\omega_e t + \frac{1}{\tau\omega_e}\sinh{\omega_e t}}), &\omega_0<\frac{1}{\tau}
    \end{cases}
\end{equation*}
where $\omega_0>1/\tau$ is called the underdamped limit, $\omega_0<1/\tau$ is called the overdamped limit, and $\omega_e=\sqrt{\omega_0^2 - 1/\tau^2}$.
With this analytical expression, it is possible to fit the intermediate scattering function $F(\boldsymbol{q},t)$ for each $\boldsymbol{q}$, with the fit parameters being the phonon frequency $\omega_0$, the damping factor $\Gamma$ (or lifetime $\tau$), and the \gls{dho} amplitude $A$.
Note that if multiple modes overlap, a sum of \glspl{dho} should be fitted, with one term for each phonon.

The fit could just as well be performed in frequency-space by Fourier transforming $F^\text{DHO}(t)$, yielding
\begin{equation*}
    S^\text{DHO}(\omega) = A\frac{2\Gamma\omega_0^2}{(\omega^2 - \omega_0^2)^2+(\Gamma \omega)^2},
\end{equation*}
and fitting the dynamic structure factor $S(\boldsymbol{q}, \omega)$ for a particular $\boldsymbol{q}$.  
Due to the density and velocity \glspl{acf} being related through $C^\text{DHO}(t) = -\ddot{F}^\text{DHO}(t)$, or, equivalently, $C^\text{DHO}(\omega) = -\omega^2{S}^\text{DHO}(\omega)$ the analytical expressions can be used to fit the calculated current correlations as well.

On a final note, let us return to the assumption of the stochastic force $f(t)$ being white noise. 
The stochastic force can be viewed as a model of the interaction between the mode being studied and other modes, so choosing it to be white noise means that the thermal bath is assumed to consist of a continuum of modes spanning all frequencies, which interact equally with the mode of interest. 
Deviations from the \gls{dho} model imply more complex dynamics, which can show up as extra peaks or wide backgrounds in the computed correlation functions. 
This can suitably be analyzed by subtracting the \gls{dho} spectrum from the corresponding structure factor or current correlation.
Remaining features are signs of, e.g., hybridization or other cases where the \gls{dho} model fails.

\subsubsection{Spherical Averaging in q}

There are several cases where it is advantageous to spherically average the computed correlation functions. 
When probing static structure factors, they are generally measured as a function of the scattering angle $2\theta$, which relates to $q=|\boldsymbol{q}|$ via
\begin{equation}\label{eq:q-to-theta}
    \sin\theta = \frac{q \lambda}{4\pi},
\end{equation}
where $\lambda$ is the wavelength of the probe.
When comparing the static structure factor $S(\boldsymbol{q})$ computed with \dynasor{} to experimentally measured ones for powder samples, we thus want to perform a spherical average in $\boldsymbol{q}$-space to obtain $S(q)$.
Furthermore, when reporting dynamic structure factors for liquids or amorphous materials these are also often visualized as a function of $q$, as the path through high-symmetry $q$-points used in crystalline solids does not extend to these systems.

\section{Examples}

\begin{figure*}
    \centering
    \includegraphics{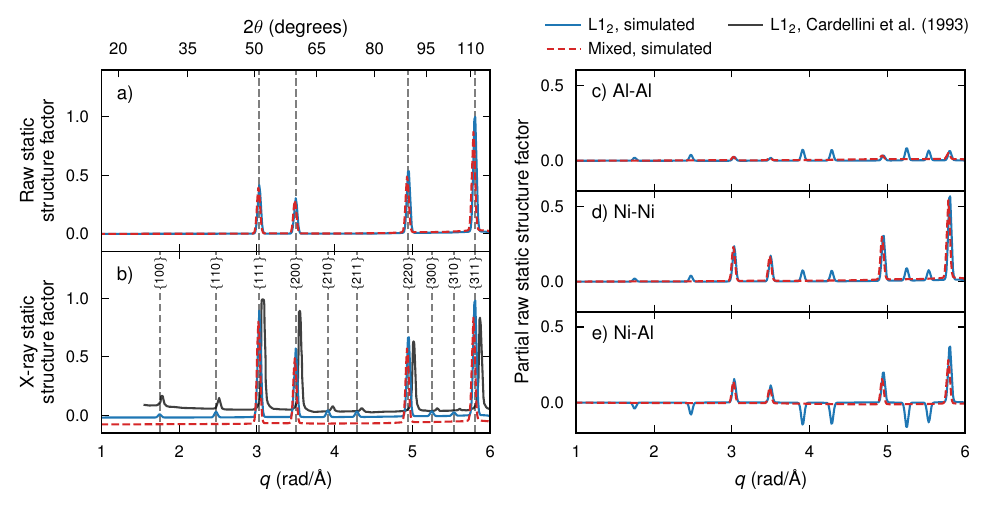} 
    \caption{
    (a) Spherically averaged raw static structure factor, $S(q)$, calculated with \dynasor{} for L$1_2$ and mixed \ce{Ni3Al}.
    The dashed lines indicate the allowed peak positions in a face-centered cubic lattice (all Miller indices even or all odd).
    (b) X-ray weighted spherically averaged static structure factors for the L$1_2$ and mixed phases, together with \gls{xrd} data extracted from Ref.~\citenum{CarCleMaz93}.
    The dashed vertical lines indicate the allowed peak positions in a simple cubic lattice.
    The curves are slightly offset to increase visibility of the smallest peaks.
    (c--e) Partial raw static structure factors for Al-Al (c), Ni-Ni (d) and Ni-Al (e) in the L$1_2$ and mixed phases.
    }
    \label{fig:NiAl_Sq}
\end{figure*}

\subsection{Crystalline and Liquid \texorpdfstring{\ce{Ni3Al}}{Ni3Al}}

In this initial example, we showcase basic usage of \dynasor{} by studying the structure and dynamics of three different phases of a nickel aluminum alloy (\ce{Ni3Al}).
These are the L$1_2$, randomly mixed, and liquid phases, respectively, which allow for the illustration of differences and similarities between working with ordered and disordered crystalline as well as liquid systems.
Furthermore, we demonstrate the weighting of computed correlation functions with X-ray form factors to allow for comparison to \gls{xrd} experiment.

First, however, the details of the underlying \gls{md} simulations must be discussed. 
There are some important considerations when generating \gls{md} trajectories for subsequent use as \dynasor{} input.
In particular, the frequency of writing atomic configurations to file (also often referred to as dump frequency), trajectory length, and simulation cell size are of importance. 

The dump frequency (together with the \gls{md} time step) sets the highest resolvable frequency. 
This means that if there are, e.g., hydrogen atoms in the system, the dump frequency must be much higher than when there are only heavier atoms present, to resolve the highest frequency as dictated by Nyqvist's theorem and to avoid FFT aliasing artifacts in the spectra. 
Here, for \ce{Ni3Al}, a dump frequency of \qty{5}{fs} is used.

The trajectory length (together with the window size in \dynasor{}), on the other hand, sets the limit for the lowest resolvable frequency. 
For this example, trajectories of \qty{1}{\nano\second} were generated using \gpumd{} \cite{FanWanYin22}, with the \gls{mlip} of the \gls{nep} form from Ref.~\citenum{fan_2023_unep}. 
For each phase, 20 separate trajectories were generated, to improve statistics by averaging over the trajectories. 
In principle, this is the same as generating one \qty{20}{\nano\second} trajectory and employing a \dynasor{} window of \qty{1}{\nano\second} with no overlap.
Using multiple shorter trajectories is, however, preferable from a computational point of view, as we do not need to resolve any lower frequencies than the one set by the \qty{1}{\nano\second} limit here, and the sampling of the potential energy surface can be improved through multiple initial conditions. 

The cell size sets the limit for smallest available $\boldsymbol{q}$. In this example, cells with side lengths of \qtyrange{86}{92}{\angstrom} (containing \qty{55296} atoms) were used, giving good resolution even at small $\boldsymbol{q}$.
In general, these \gls{md} parameters should be carefully considered and converged. 

\subsubsection*{Structure: X-ray Diffraction}

We begin by studying the structure of the two solid (\ce{Ni3Al}) phases as represented by the static structure factor, $S(q)$ (\autoref{fig:NiAl_Sq}a).
When studying structure, we do not have to bother with the considerations of sufficiently high dump frequency in the \gls{md}, as discussed in the previous section, because there is no time-dependence in the correlation function.
Therefore, 1000 frames from each of the 20 trajectories are used, with \qty{25}{\femto\second} between each frame, to give a representative view of the structure of the material and obtain nicely converged static structure factors.
The dashed vertical lines in \autoref{fig:NiAl_Sq}a indicate the expected position of the Bragg peaks for a \gls{fcc} lattice, i.e., where all Miller indices are even or all are odd, with a lattice constant of \qty{3.5915}{\angstrom}. 
Both the L$1_2$ peaks and the mixed phase peaks arise exactly at these $q$-values, as there is no atom type-specific information in the raw structure factor calculation distinguishing the \ce{Ni} from the \ce{Al} atoms, so it is only the underlying \gls{fcc} lattice that dictates the result. 
The relative peak intensities are as expected based on the multiplicities of each of these Bragg peaks: $8\times \{111\}$, $6\times \{200\}$, $12\times \{220\}$, and $24\times \{311\}$.
The computed raw static structure factor thus contains information about all resolved $q$-values.

Now to the atom type-specific information, which in this case is introduced by weighting the static structure factor with X-ray form factors.
These form factors are $q$-dependent and vary with atom type, as they scale with the number of electrons. 
When the static structure factors are weighted, we see a difference arise between the mixed and the L$1_2$ phase, as the latter exhibits additional peaks (\autoref{fig:NiAl_Sq}b).
These emerging peaks match those present in an \gls{xrd} measurement of crystalline \ce{Ni3Al}, illustrating good agreement between X-ray form factor weighted $S(q)$ and experiment.
The peak positions for the experiment are slightly offset compared to the calculated ones, owing to a minor difference in lattice constant between computation and experiment.
Differences in peak heights indicate that the measured system is slightly off-stoichiometric, compared to the simulated ideal L$1_2$ structure.

The total weighted L$1_2$ structure factor exhibits peaks at the positions expected from a \gls{sc} system, while the mixed phase only exhibits the \gls{fcc} peaks. 
This can be understood by investigating the unweighted partial structure factors, readily obtained from the \dynasor{} calculation (\autoref{fig:NiAl_Sq}c--e). 
The L$1_2$ aluminum sublattice forms a \gls{sc} structure with the same lattice constant, as each unit cell only contains one \ce{Al} atom, so the partial Al--Al structure factor (\autoref{fig:NiAl_Sq}c) has peaks according to such a \gls{sc} lattice.
The randomly mixed system does not exhibit this sublattice order, but overall the atoms still occupy a \gls{fcc} lattice, so the \gls{fcc} peaks are visible.
The same explanation holds for the Ni--Ni partial structure factor, though an interesting observation can be made about the Ni--Al partial (\autoref{fig:NiAl_Sq}e).
Here, we see that the non-\gls{fcc} peaks are negative for the L$1_2$ phase, exactly canceling the corresponding peaks in the other partials when summed, such that the raw total structure factor only exhibits the expected \gls{fcc} peaks (\autoref{fig:NiAl_Sq}a).
When weighting with X-ray form factors, however, these three partials are weighted differently, such that the negative peaks in the Ni--Al partial no longer cancel the positive peaks in the other two, leading to a result that matches experiment (\autoref{fig:NiAl_Sq}b).
Due to the type of each atom being stored during \gls{md}, and the partial correlations being readily available from \dynasor{}, we can thus know which parts of the system contribute to each peak.

\subsubsection*{Dynamics: Current Correlations}

\begin{figure*}
    \centering
    \includegraphics{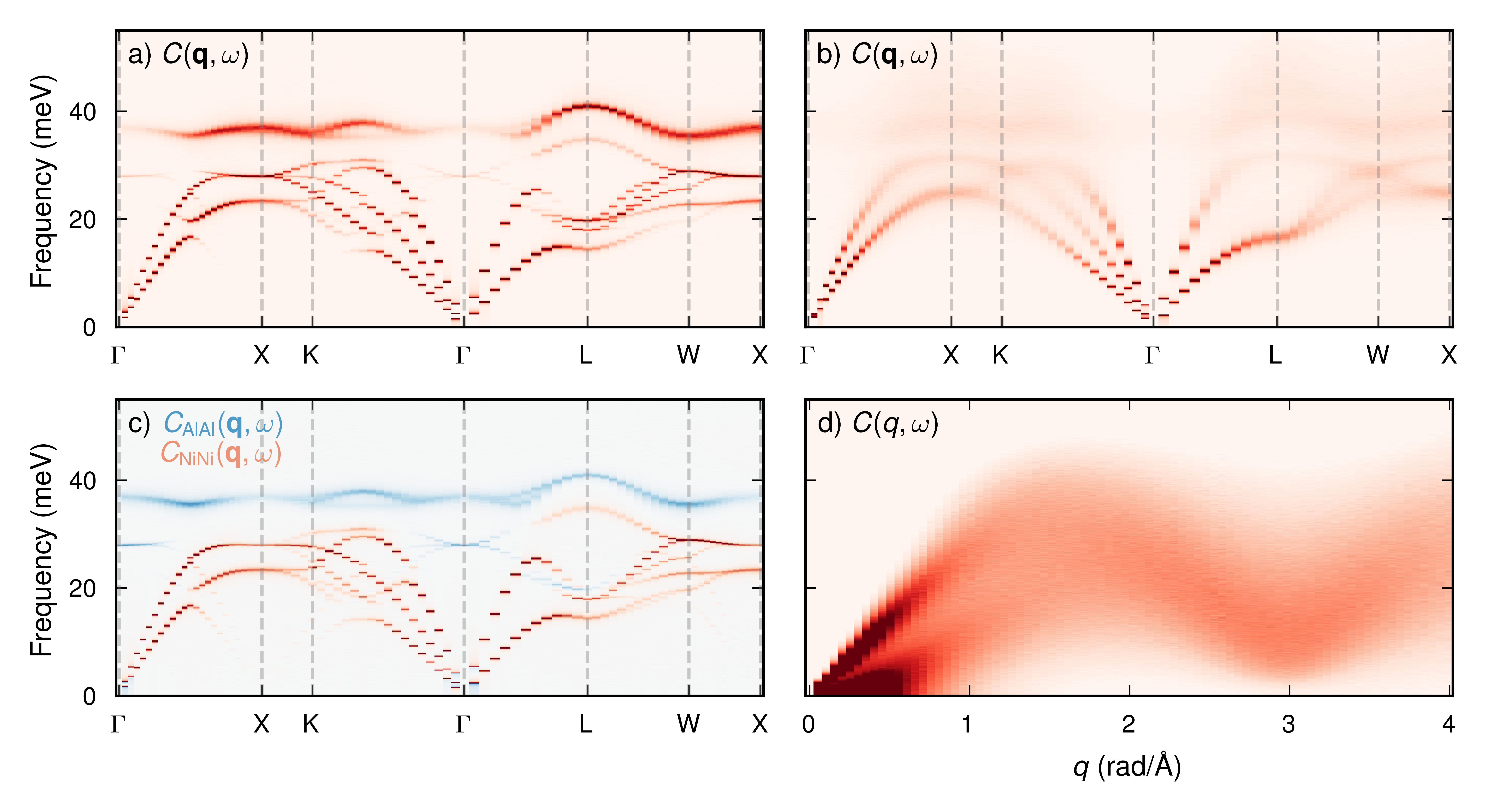}
    \caption{Current correlation $C(\boldsymbol{q},\omega)$ for (a) ordered crystalline \ce{Ni3Al}, (b) disordered crystalline \ce{Ni3Al}, and (d) liquid \ce{Ni3Al}, as well as (c) partial current correlations (Ni--Ni in red and Al--Al in blue) of ordered \ce{Ni3Al}.
    Note that a face-centered cubic high-symmetry path is used in (a--c).
    }
    \label{fig:NiAl_Cqw}
\end{figure*}

To learn about the dynamics of the three different phases of the \ce{Ni3Al} alloy, we study the current correlations, $C(\boldsymbol{q}, \omega)$. 
As mentioned in \autoref{sec:theory-correlation-functions}, current correlations can be viewed as spatially dependent generalizations of the velocity \gls{acf}, which means that mode-specific vibrational frequencies can be obtained by computing $C(\boldsymbol{q}, \omega)$. 
The phonon dispersion for the L$1_2$ phase of \ce{Ni3Al} (\autoref{fig:NiAl_Cqw}a) is computed along a \gls{fcc} high-symmetry path, which reveals backfolding, e.g., along $\Gamma$--X, due to a doubling of the unit cell.  
This dispersion, based on current correlations, includes both longitudinal and transverse modes, in contrast to computed dynamic structure factors,  which would only exhibit the longitudinal modes following Eq.~\eqref{eq:relation-S-Cl}.  

Comparing ordered (\autoref{fig:NiAl_Cqw}a) and disordered crystalline (\autoref{fig:NiAl_Cqw}b) \ce{Ni3Al}, we can furthermore note that the disorder causes a smearing of the sharp features in the dispersion of the L$1_2$ phase. 
Additionally, with this approach to computing phonon dispersions, we also have access to the dispersion relation for the liquid phase (\autoref{fig:NiAl_Cqw}d), which is something that would be unobtainable with perturbative methods. 
At first glance, the liquid dispersion might look very different from those for the solid phases (\autoref{fig:NiAl_Cqw}ab).
This is, however, due to $C(\boldsymbol{q}, \omega)$ being spherically averaged in $\boldsymbol{q}$ for the liquid, owing to its isotropic nature.
With this in mind, we can recognize that the liquid dispersion is an even more smeared out version of the solid dispersion, see, e.g., the similarity between the liquid dispersion and the path between $\Gamma$ and $X$ (\autoref{fig:NiAl_Cqw}b).
On a more technical note, we can observe that there is no intensity whatsoever close to $q=0$ in the liquid dispersion, which is due to the finite size of the \gls{md} simulation posing a limit on the smallest available $\boldsymbol{q}$.

Returning to the current correlation functions for the crystalline phases, there is, in fact, even more insight to be gained from these simulations. 
As for the static structure factor, partials are readily available for dynamic structure factors and current correlations alike. 
Examples of partial current correlations for the L$1_2$ phase of \ce{Ni3Al} are shown in \autoref{fig:NiAl_Cqw}c. 
This immediately gives us information about which contributions stem from Ni--Ni interactions, and which stem from Al--Al interactions, something that can be cumbersome to discern purely from experimental spectra, especially for systems with even more atomic species.

\subsection{\texorpdfstring{\ce{BaZrO3}}{BaZrO3}}

In this example, we study the cubic oxide perovskite \ce{BaZrO3} using the \gls{mlip} potential based on the \gls{nep} formalism developed in Ref.~\citenum{FraRosErhWah2023}.

\ce{BaZrO3} has attracted much interest as it retains its cubic structure down to \qty{0}{\kelvin} at ambient pressure \cite{Akbarzadeh2005, Knight2020, PerJedRomPioLinHylWah20}, while most perovskites have a non-cubic low temperature phase.
However, it is clear that the antiferrodistortive out of phase tilting mode softens substantially with decreasing temperature \cite{Zheng2022, RosFraBrauTouBouAndBosMaeWah23}, giving rise to interesting local correlations \cite{Levin2021, FraRosErhWah2023}.
To aid us in studying this material, we weight the computed correlation functions with electron form factors, allowing for comparison with electron diffraction measurements, and exploit mode projection to gain more insight into the microscopic dynamics giving rise to the unique low-temperature behavior of \ce{BaZrO3}.

\subsubsection*{Structure: Electron Diffraction}

\begin{figure}
    \centering
    \includegraphics{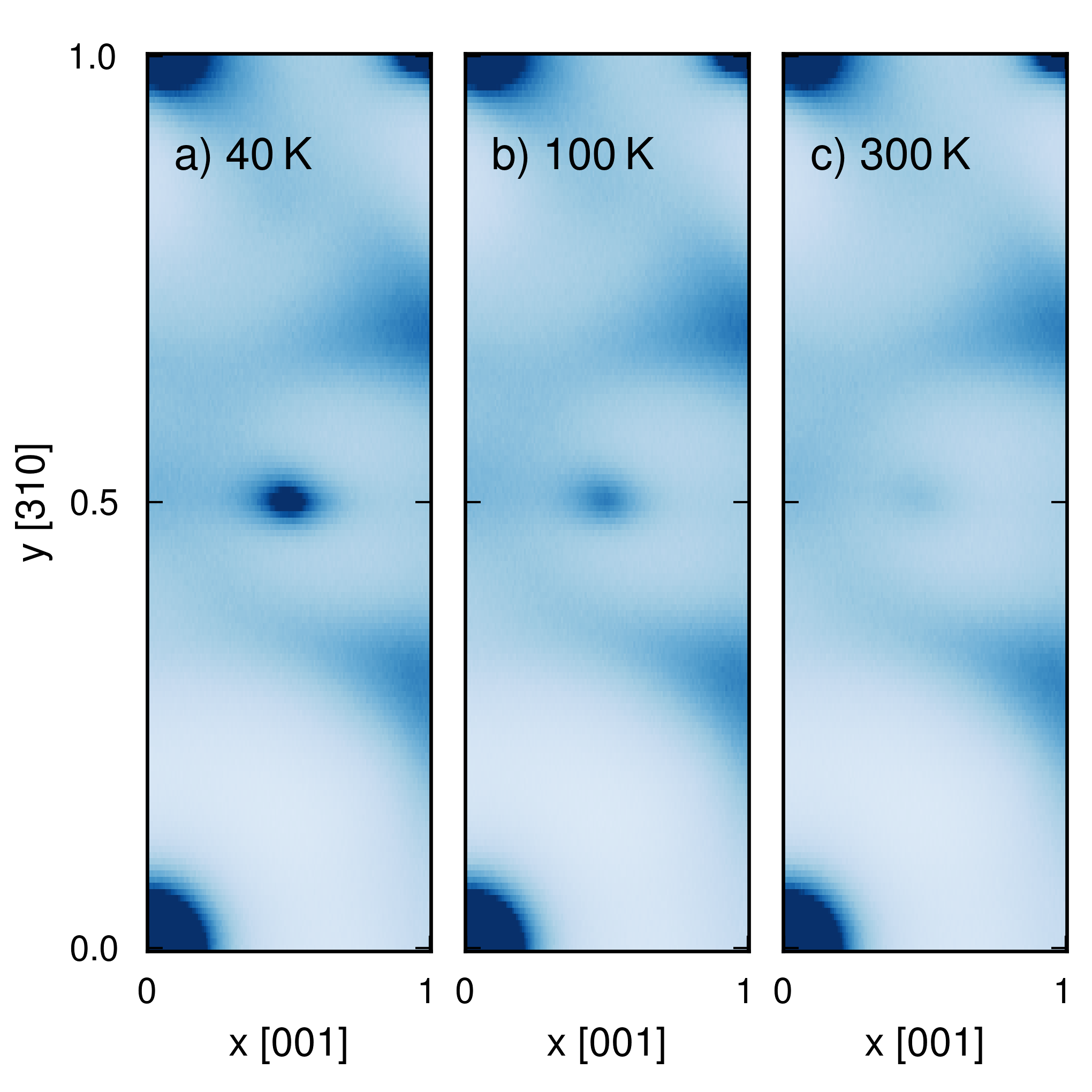}
    \caption{
    Electron beam diffraction intensity, $S(\boldsymbol{q})$, calculated from \gls{md} simulations at a) \qty{40}{\kelvin}, b) \qty{100}{\kelvin}, and c) \qty{300}{\kelvin}, reproduced from Ref.~\citenum{FraRosErhWah2023}.
    Here, $S(\boldsymbol{q})$ are evaluated for $\boldsymbol{q}$-points given by $\boldsymbol{q} = \frac{2\pi}{a}(3y, y, x)$.
    The center of the heatmaps corresponds to the R-point, i.e. $\boldsymbol{q}=\frac{2\pi}{a}(3/2,1/2,1/2)$.
    }
    \label{fig:BZO_Sq}
\end{figure}

\begin{figure*}
    \centering
    \includegraphics[scale=0.254]{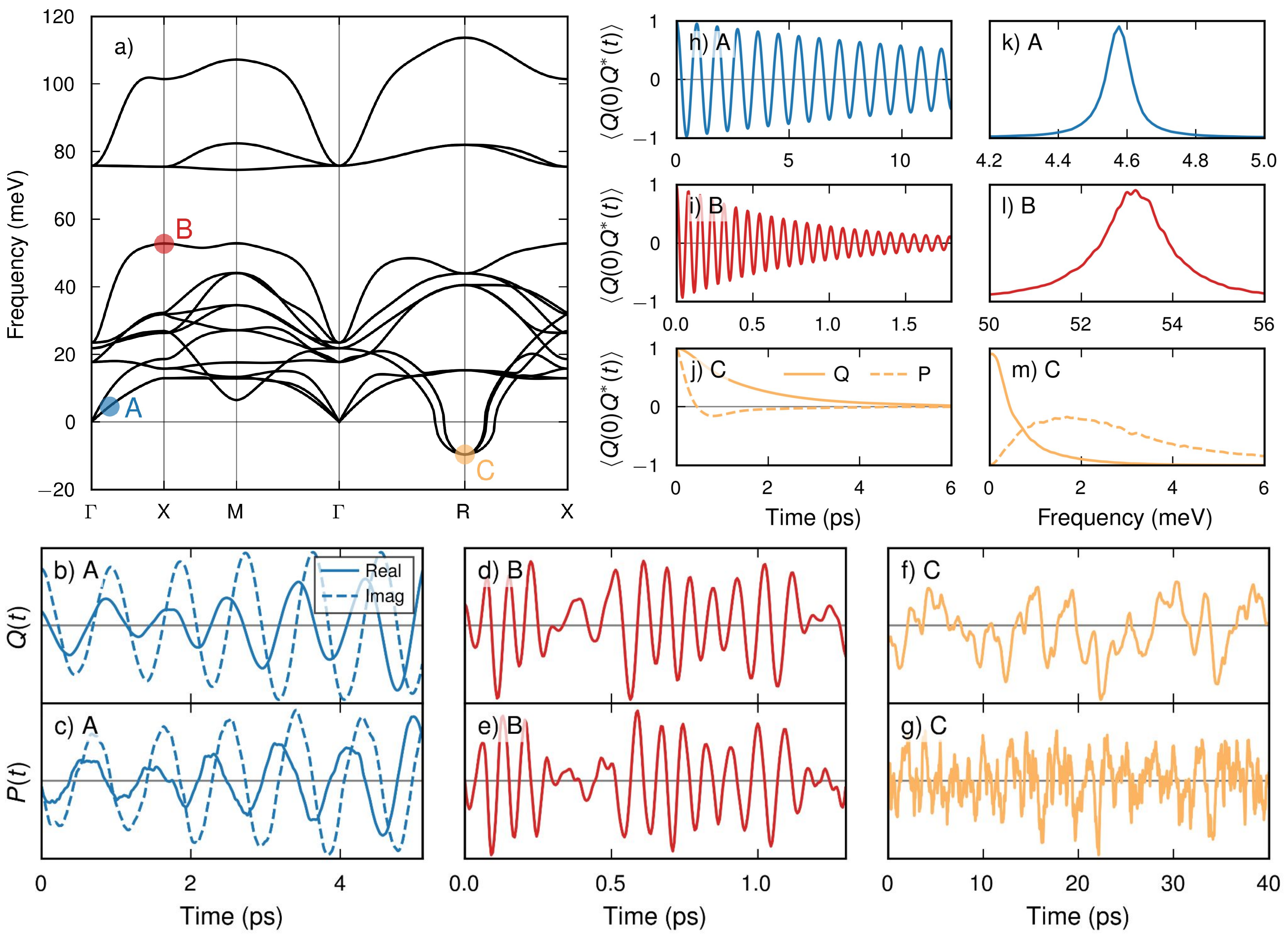}
    \caption{
    Mode projection analysis in \ce{BaZrO3} at \qty{300}{\kelvin} and \qty{15.6}{\giga\pascal}.
    Harmonic phonon dispersion (a) computed with \phonopy{} at the lattice-parameter corresponding to \qty{15.6}{\giga\pascal} at \qty{300}{\kelvin}.
    Here, the selected modes for phonon mode projection are highlighted in blue (A), red (B), and orange (C).
    The mode coordinates $Q(t)$ (b,d,f) and momenta $P(t)$ (c,e,g) are shown as a function of time.
    For the acoustic mode (A) the eigenvector is imaginary and consequently $Q$ and $P$ are as well, whereas the two zone-boundary modes (B, C) are real.
    The mode \glspl{acf} $\left < Q(0) Q^*(t) \right >$ are shown in (h, i, j) and their Fourier transforms in (k, l, m).
    For the A and B modes the \glspl{acf} are hardly distinguishable between $Q$ and $P$, hence the latter are not shown, whereas they differ significantly for the overdamped mode C.
    The spectrum for $P$ in (m) is scaled with an arbitrary factor to be comparable to the spectrum of $Q$. 
    }
    \label{fig:BZO_mode_projection}
\end{figure*}

To gain insight into the temperature-dependence of the structure of \ce{BaZrO3}, diffuse scattering due to phonons can be studied by monitoring the static structure factor, $S(\boldsymbol{q})$, between the Bragg peaks.
In Ref.~\citenum{Levin2021} diffuse scattering was measured by electron beam diffraction in a 2D space spanned by q-vectors $\left<310\right>$ and $\left<001\right>$ such that an R-point ($\left<1.5, 0.5, 0.5\right>$) sits in the middle.
This diffuse scattering was simulated in Ref.~\citenum{FraRosErhWah2023} and in \autoref{fig:BZO_Sq} we reproduce the resulting electron form factor-weighted $S(\boldsymbol{q})$ for three temperatures: 40, 100, and \qty{300}{\kelvin}.
The simulated structure factors clearly show the emergence of an intense broad peak at the R-point as temperature decreases, which is in good agreement with experimental results from Levin \emph{et al.} \cite{Levin2021}.
With the aid of these simulations, it is understood that the emerging diffuse peak arises from the soft octahedral tilt mode located at R due to the frequency of this mode softening substantially as temperature decreases \cite{FraRosErhWah2023}.

\subsubsection*{Dynamics: Mode Projection}

At room temperature, with the present \gls{mlip}, \ce{BaZrO3} undergoes a phase transition from the cubic to the tetragonal structure for pressures around \qty{16}{\giga\pascal}, driven by the soft tilt mode at the R-point \cite{FraRosErhWah2023}.
This means that the mode exhibits overdamped dynamics in the vicinity of the transition.
In order to study this further, we employ mode projection at \qty{300}{\kelvin} and \qty{15.6}{\giga\pascal} for the cubic phase, which requires harmonic force constants and \gls{md} trajectories (\autoref{fig:workflow}).
Additionally, we contrast the behavior of this overdamped mode by analyzing two other normal modes.

First, the harmonic force constants are calculated at the lattice-parameter corresponding to \qty{15.6}{\giga\pascal} at \qty{300}{\kelvin} with \phonopy{} \cite{phonopy-phono3py-JPSJ, phonopy-phono3py-JPCM}.
These are used to construct the phonon dispersion and the phonon eigenvectors, visualized in \autoref{fig:BZO_mode_projection}a where modes selected for the mode projection are highlighted.
Note that while the R-point mode is imaginary at \qty{0}{\kelvin} it is dynamically stabilized at conditions (\qty{300}{\kelvin} and \qty{15.6}{\giga\pascal}) considered here.
\Gls{md} simulations are carried out in a $24\times24\times24$ supercell (\qty{69120} atoms) with a timestep of \qty{1}{\femto\second} and a dump frequency of \qty{5}{\femto\second}.
Mode projection \glspl{acf} and power spectra are averaged over about \qty{50}{\nano\second} in total.

The phonon mode coordinate $Q(t)$ and momenta $P(t)$ are obtained via mode projection from the \gls{md} simulations (\autoref{fig:BZO_mode_projection}b--g).
Because the acoustic mode (A) is the only selected mode that has an imaginary eigenmode $X_{sb}^\alpha(\boldsymbol{l}, \boldsymbol{q})$, only this mode has imaginary $Q(t)$ and $P(t)$ (\autoref{fig:BZO_mode_projection}b,c).
For mode A we see how the real and imaginary part oscillate much like a harmonic system, and the momentum $P$ is lagging behind the oscillation of coordinate $Q$ by about 1/4 of the period as expected for a harmonic oscillator.
The oscillations also appear mostly harmonic for mode B (\autoref{fig:BZO_mode_projection}d,e), but with some noise or irregularities, which indicates anharmonic features.
Lastly, mode C (\autoref{fig:BZO_mode_projection}f,g) does not follow a clear oscillating pattern but rather exhibits the dynamics of an overdamped mode \cite{FraRosEriRahTadErh23}.

To analyze the mode coordinates in a statistical manner we compute their \glspl{acf}, $\left < Q(0) Q^*(t) \right > / \left < Q(0) Q^*(0) \right >$ (\autoref{fig:BZO_mode_projection}h--j), and the corresponding Fourier transforms (\autoref{fig:BZO_mode_projection}k--m).
The \gls{acf} for mode A oscillates and decays very slowly, indicating a long phonon lifetime. 
This is also seen as a sharp peak in the frequency domain.
For mode B the \gls{acf} decays to zero after about \qty{2}{\pico\second}, giving rise to a broader peak in the frequency domain.
Note that the \gls{acf} and spectra of $Q$ and $P$ are almost identical for both mode A and mode B, as is expected for weakly anharmonic modes, where $\omega_0 \tau \gg 1$.
Mode C, however, behaves differently, as the \gls{acf} for $Q$ decays towards zero instead of oscillating, due to the mode being overdamped, i.e., $\omega_0 \tau < 1$.
The \gls{acf} for $P$ also deviates from that of $Q$, as is expected when the lifetime $\tau$ is similar to (or smaller than) one time period of the oscillation with angular frequency $\omega_0$.
Additionally, the Fourier transform of $Q$ does not exhibit a peak at a non-zero frequency, but rather a peak centered around zero.
This is typical behavior of an overdamped mode, where $\omega_0 \tau < 1$.
While this mode is very anharmonic, it is worthwhile to note that it is still almost perfectly described by the \gls{dho} model in \autoref{sec:post-processing}.

\begin{figure*}
    \centering
    \includegraphics{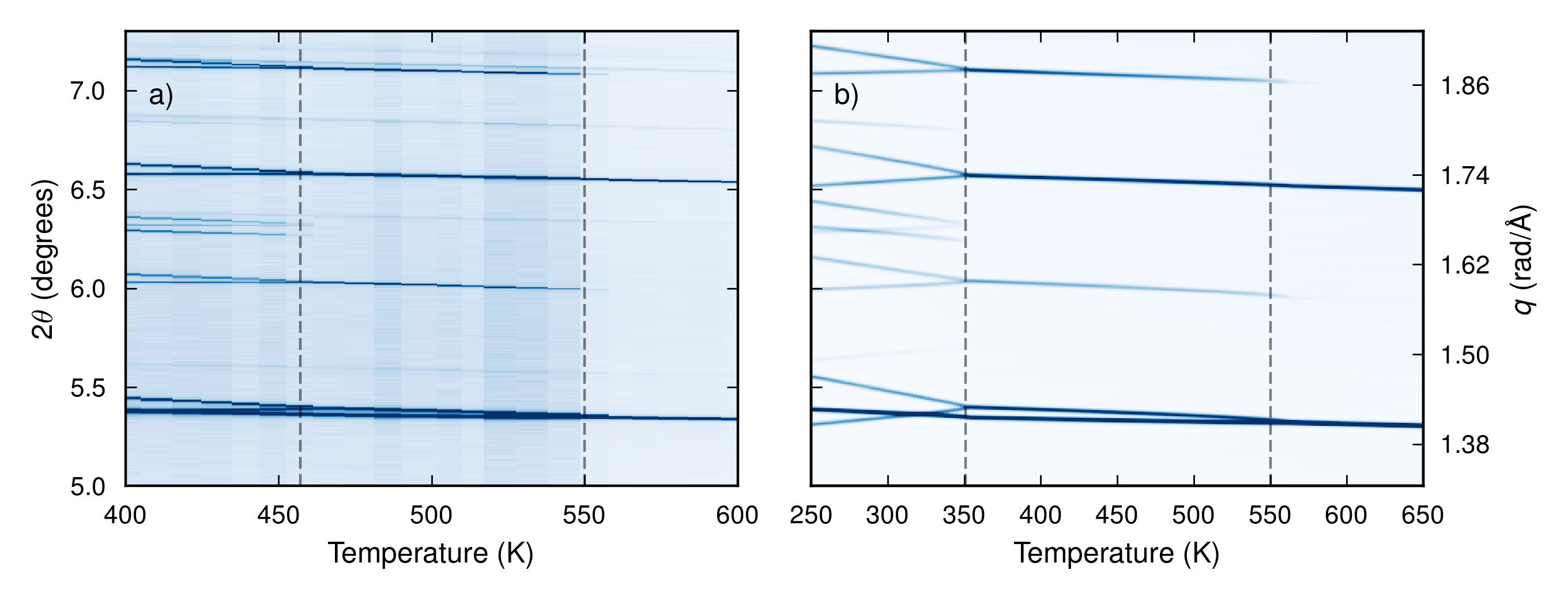}
    \caption{
        Heatmaps of (a) \gls{xrd} intensities from experimental measurements \cite{Marronnier2018} and (b) simulated, X-ray weighted, spherically averaged static structure factor, $S(q)$, for \ce{CsPbI3} as a function of temperature.
        Dashed vertical lines indicate the phase transition temperatures.
        The experimental and simulated spectra are scaled with arbitrary factors to align the peak intensities.
    }
    \label{fig:CsPbI3_Sq}
\end{figure*}

An additional benefit of phonon mode projection is that isolation of degrees of freedom with different time-scales can allow for improved sampling.
For example, a low-frequency phonon mode with a long lifetime (many nanoseconds) requires very long \gls{md} simulations to sample properly, and if the material at hand contains light atoms such as hydrogen, the trajectory must be sampled approximately every \qty{5}{\femto\second} to avoid aliasing when computing other correlation functions such as \gls{sed} or dynamic structure factors.
However, if the dynamics is projected onto the low-frequency mode of interest, the analysis can be carried out with a much lower \gls{md} dump frequency, because the fast vibrations are not present in the relevant time signals for the low-frequency mode.

Beyond the analysis carried out for these three modes in \ce{BaZrO3}, phonon mode projection can be employed to understand detailed contributions to vibrational spectra, e.g., in Raman spectra \cite{Rosander2024}, or the heat transfer and correlation between modes \cite{Rohskopf2022}.
Having access to all the phonon modes and easily being able to manipulate them in atomic-scale simulations also allows for conducting non-equilibrium simulations where for example a few select modes are heated or cooled by, e.g., an external driving force or a thermostat.
Therefore, mode projection is an excellent method for studying detailed phonon dynamics in any crystalline system using \gls{md} simulations in combination with \dynasor{}.

\subsection{\texorpdfstring{\ce{CsPbI3}}{CsPbI3}}

In the final example, we analyze the prototypical halide perovskite \ce{CsPbI3}, using the \gls{mlip} in \gls{nep} format from Ref.~\citenum{FraWikErh2023} to conduct the underlying \gls{md} simulations.

The ground state of \ce{CsPbI3} is a non-perovskite phase, often referred to as the $\delta$- or needle-like phase, with space group \hmn{Pnma}.
Upon heating, the material undergoes a transition to a cubic perovskite phase (\hmn{Pm-3m}) around \qty{600}{\kelvin}.
When cooling the material, it first undergoes a transition to a tetragonal phase (\hmn{I4/mcm}) around \qty{540}{\kelvin} and then to an orthorhombic perovskite phase (\hmn{Pnma}) around \qty{430}{\kelvin} \cite{Marronnier2018}.
To gain insight into the atomic mechanisms occurring in \ce{CsPbI3} when undergoing these phase transitions, computed correlation functions are weighted with X-ray form factors and neutron scattering lengths, and compared to relevant experiments, in order to study the structure and dynamics, respectively.
Additional insight into the dynamics is provided via \gls{sed} calculations.

\subsubsection*{Structure: X-ray Diffraction}

We begin by showing how phase transitions within the perovskite can be identified by computing and analyzing the static structure factor $S(q)$, as done in Ref.~\citenum{Kayastha2025} for another perovskite.
$S(q)$ is computed for all $q< \qty{2.2}{\rad\per\angstrom}$ that are commensurate with the supercell, which consists of approximately \qty{40000} atoms.
For each temperature $S(q)$ was computed from 100 snapshots, evenly spaced by \qty{1}{\pico\second}.

The simulated diffraction data closely resemble the data from Ref.~\citenum{Marronnier2018} in terms of additional peaks and peak splitting at the phase transitions (\autoref{fig:CsPbI3_Sq}), although the lower phase transition temperature is about \qty{100}{\kelvin} lower with this \gls{mlip} compared to the experiment, a feature that is intrinsic to the exchange-correlation functional used for underlying reference calculations \cite{FraWikErh2023}.

Here, we convert $q$-points to the scattering angle $2\theta$ used in the experiment through Eq.~\eqref{eq:q-to-theta} with the X-ray wavelength $\lambda=\qty{0.413906}{\angstrom}$ from Ref.~\citenum{Marronnier2018}.

At the upper transition two additional peaks appear at $q$-values of \qty{1.58}{\rad\per\angstrom} and \qty{1.86}{\rad\per\angstrom}, corresponding to $q$-pointts $[1.5, 0.5, 0.0]$ and $[1.5, 0.5, 1.0]$ in the cubic structure. 
The reason for this is that one phonon mode in the cubic structure, located at $[0.5, 0.5, 0]$ (M-point), condenses and freezes in at this transition.
Additionally, the peak at \qty{1.4}{\rad\per\angstrom}, corresponding to $[1, 1, 0]$ in the cubic structure, splits as the system becomes tetragonal.

At the lower transition, additional modes condense giving rise to new peaks around \qty{1.65}{\rad\per\angstrom}, corresponding to $[1.5, 0.5, 0.5]$ (R-point), and at 1.5 and \qty{1.8}{\rad\per\angstrom} corresponding to $[0.5, 1, 1]$ and $[1.5, 1, 1]$ (X-points), respectively.
Existing peaks also split as a result of the structure transitioning from the tetragonal to the orthorhombic phase.
A difference between simulation and experiment is that the peak splitting is larger in the computed structure factor (\autoref{fig:CsPbI3_Sq}b), which is due to the simulations predicting larger differences between the lattice parameters in the orthorhombic phase than what is observed in the experiment, which means that peaks at, e.g., $2\pi/a$ and $2\pi/c$ are farther apart.
    
\subsubsection*{Dynamics: Inelastic Neutron Scattering}

\begin{figure*}
    \centering
    \includegraphics{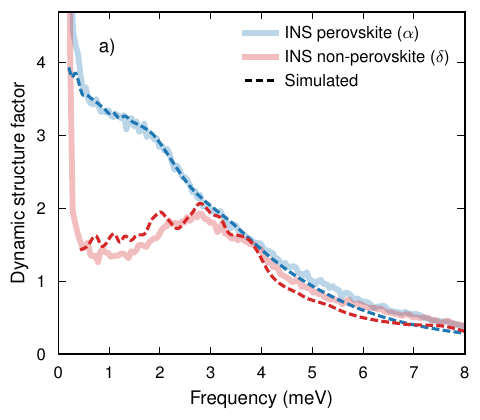}
    \includegraphics{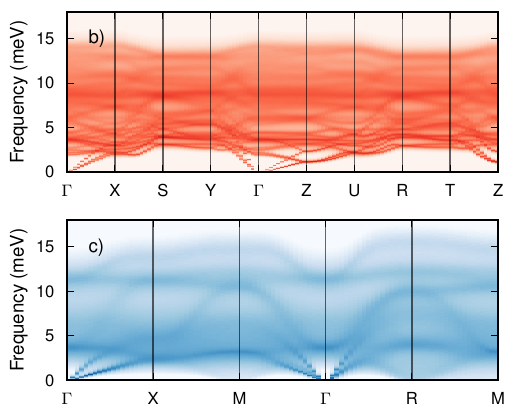}
    \caption{
        (a) \Gls{ins} measurements from Ref.~\citenum{Laven2025} and computed dynamic structure factor, $S(\boldsymbol{q}, \omega)$, for \ce{CsPbI3}.
        (b) \gls{sed} for the non-perovskite $\delta$-phase at \qty{550}{\kelvin}.
        (c) \gls{sed} for the cubic perovskite $\alpha$-phase at \qty{600}{\kelvin}.
    }
    \label{fig:CsPbI3_Sqw}
\end{figure*}

Having identified the phase transitions from simulations, we turn to studying the dynamics in \ce{CsPbI3} in the cubic perovskite $\alpha$-phase and in the non-perovskite $\delta$-phase.
\Gls{ins} measurements reveal a large discrepancy between the low frequency (0-\qty{2}{\milli\electronvolt}) behavior between these two phases (\autoref{fig:CsPbI3_Sqw}a) \cite{Laven2025}.
This discrepancy is well captured by the computed dynamic structure factor.
Here, both the measured and simulated dynamic structure factor is averaged over $q$-points in the range \qtyrange{0.2}{3.5}{\rad\per\angstrom}.
The elastic peak arising at $\omega=0$ manifests itself as a delta-function in the computed structure factor, while it gives rise to a broader peak centered around $\omega=0$ in experiments due to the finite resolution.
The simulated data therefore does not capture the intensity increase below \qty{0.5}{\milli\electronvolt}.

In order to understand the underlying reason for the differing behavior in dynamics between the $\alpha$- and the $\delta$-phases at low frequencies, we compute the \gls{sed} across the entire Brillouin zone for both phases (\autoref{fig:CsPbI3_Sqw}b,c).
Here, we see that the $\delta$-phase exhibits clearly distinguishable and sharper phonon modes with little broadening compared to the cubic perovskite, which has very broad modes.
The $\delta$-phase has only a few, and mostly acoustic, modes contributing to intensities in the low frequencies range, \qtyrange{0}{2}{\milli\electronvolt}, whereas the cubic perovskite phase shows notably large intensities for low frequencies at the R-point and M-point.
These modes at R and M are overdamped and thus the largest reason for the substantial discrepancy between the two phases in the \gls{ins} measurements (\autoref{fig:CsPbI3_Sqw}a).

\section{Conclusions}

We have demonstrated the utility of the \dynasor{} package in version 2.X  via a broad range of examples.
By means of three different phases of \ce{Ni3Al} and the two perovskites \ce{BaZrO3} and \ce{CsPbI3}, we have explored a number of possible insights into the atomic-scale dynamics that can be obtained directly from \gls{md}-based correlation functions computed with \dynasor{}.
For \ce{Ni3Al}, the raw correlation functions used for studying dynamics were current correlations, for \ce{BaZrO3} phonon mode \glspl{acf} were analyzed, and for \ce{CsPbI3} the phonon \gls{sed} was computed. 
The insights we gained from these approaches include understanding partial contributions to the phonon dispersions, being able to compare spectra for, e.g., solid and liquid phases, which would not be possible with perturbative approaches, and exploring the nature of phonon modes and the role they play during, e.g., phase transitions by employing \gls{sed} and phonon mode projection.

Furthermore, we have illustrated agreement with scattering experiments relying on X-rays, electrons and neutrons to probe these materials.
For solid \ce{Ni3Al}, this has been shown by comparing spherically averaged X-ray weighted static structure factors and powder \gls{xrd} patterns.
For \ce{BaZrO3} it was shown by comparing simulated and measured electron beam diffraction intensities.
For \ce{CsPbI3}, good agreement was seen between spherically averaged X-ray weighted static structure factors and measured \gls{xrd} patterns, as well as between neutron weighted dynamic structure factors and inelastic neutron scattering experiments. 
This has given us insight into the process of studying structure and dynamics using \gls{md}-based correlation functions, which allows for identification of phase transitions as the correlation functions can be computed as a function of temperature. 
In addition, the reason for the differences between the \gls{ins} spectra of two phases of \ce{CsPbI3} was explained using the simulated, neutron weighted dynamic structure factor in combination with the \gls{sed}.

It would be possible to take the comparison to experiment one step further, beyond weighting with probe-specific weights, and fully include the details of the measurement setup by convoluting the computed structure factors with resolution functions. 
This would allow for quantitative predictions of instrument-specific spectra, which could further aid both interpretation and planning of experiments, and extending \dynasor{} in this direction is actively pursued.

In conclusion, \gls{md}-based correlation functions allow for the study of intricate details of structural and dynamical material properties, including effects of temperature, for all types of condensed materials, ranging from crystalline to liquid.
\dynasor{} enables such analysis by enabling the computation and post-processing of these correlation functions, with weights readily available for the three probes, facilitating direct comparison to experiment.

\section*{Acknowledgments}

This work was funded by the Swedish Research Council (grant numbers 2020-04935, 2021-05072), the Swedish Foundation for Strategic Research via the SwedNESS graduate school (GSn15-0008), and the Chalmers Initiative for Advancement of Neutron and Synchrotron Techniques.
The computations were enabled by resources provided by the National Academic Infrastructure for Supercomputing in Sweden (NAISS) at C3SE, NSC, and PDC partially funded by the Swedish Research Council through grant agreements no. 2022-06725 and no. 2018-05973 as well as the Berzelius resource provided by the Knut and Alice Wallenberg Foundation at NSC.

We thank Rasmus Lavén and Maths Karlsson for helpful discussions and for the \gls{ins} data for \ce{CsPbI3} and the authors of Ref.~\citenum{Marronnier2018} for providing their \gls{xrd} data for \ce{CsPbI3}.

\end{document}